\documentclass[aps,pra,twocolumn,amsmath,showpacs,superscriptaddress]{revtex4}

\usepackage{graphicx}
\usepackage{amssymb}
\usepackage{amsmath}

\def\mathbi#1{\textbf{\em #1}}

\begin{document}

\title{Controlled deflection of cold atomic clouds and of Bose-Einstein condensates}

\author{N. Gaaloul}

\affiliation{Laboratoire de Spectroscopie Atomique, Mol\'{e}culaire et Applications,
D\'{e}partement de Physique, Facult\'{e} des Sciences de Tunis,
Universit\'{e} Tunis El Manar, 2092 Tunis, Tunisia.}

\affiliation{Laboratoire de Photophysique Mol\'{e}culaire du CNRS,
Universit\'{e} Paris-Sud, B\^{a}timent 210,
91405 Orsay Cedex, France.}

\author{A. Jaouadi}

\affiliation{Laboratoire de Spectroscopie Atomique, Mol\'{e}culaire et Applications,
D\'{e}partement de Physique, Facult\'{e} des Sciences de Tunis,
Universit\'{e} Tunis El Manar, 2092 Tunis, Tunisia.}

\affiliation{Laboratoire de Photophysique Mol\'{e}culaire du CNRS,
Universit\'{e} Paris-Sud, B\^{a}timent 210,
91405 Orsay Cedex, France.}

\author{L. Pruvost}

\affiliation{Laboratoire Aim\'{e} Cotton du CNRS,
Universit\'{e} Paris-Sud, B\^{a}timent 505,
91405 Orsay Cedex, France.}

\author{M. Telmini}

\affiliation{Laboratoire de Spectroscopie Atomique, Mol\'{e}culaire et Applications,
D\'{e}partement de Physique, Facult\'{e} des Sciences de Tunis,
Universit\'{e} Tunis El Manar, 2092 Tunis, Tunisia.}

\author{E. Charron}

\affiliation{Laboratoire de Photophysique Mol\'{e}culaire du CNRS,
Universit\'{e} Paris-Sud, B\^{a}timent 210,
91405 Orsay Cedex, France.}

\abstract{We present a detailed, realistic proposal and analysis of the implementation of a cold atom deflector using time-dependent far off-resonance optical guides. An analytical model and numerical simulations are used to illustrate its characteristics when applied to both non-degenerate atomic ensembles and to Bose-Einstein condensates. Using for all relevant parameters values that are achieved with present technology, we show that it is possible to deflect almost entirely an ensemble of $^{87}$Rb atoms falling in the gravity field. We discuss the limits of this proposal, and illustrate its robustness against non-adiabatic transitions.}

}

\maketitle

\section{Introduction}
\label{sec:Intro}

Optical and magnetic fields are extremely efficient tools used for the controlled manipulation of large ensembles of cold atoms\,\cite{Adams_1994,Balykin_1995}. In the past fifteen years, cold matter waves have shown great possibilities in the context of linear atom optics, when phase-space densities are sufficiently low that the effect of collisions can be neglected. Dipole and radiation-pressure forces have for instance allowed the achievement of various optical manipulations such as atomic focusing, diffraction or interference\,\cite{Berman_1997,Meystre_2001}.

Many efforts have been recently devoted to the experimental implementation of atomic beam splitters with magnetic\,\cite{Muller_2000,Cassettari_2000,Muller_2001,Hommelhoff_2005} or optical\,\cite{Houde_2000,Hansel_2001,Dumke_2002} potentials. These different experimental investigations were accompanied by various theoretical studies\,\cite{Stickney_2003,Kreutzmann_2004,Bortolotti_2004,Gaaloul_2006,Zhang_2006}. These devices are obviously of clear interest for atom interferometry experiments. After the advent of Bose-Einstein condensation (BEC) in 1995\,\cite{Anderson_1995,Davis_1995}, different setups were designed in order to split and recombine a BEC\,\cite{Shin_2004,Wang_2005,Schumm_2005}. In this case, the experimental implementation is even more difficult since inter-atomic interactions due to high atomic densities in the wave-guides can sometimes not only induce the fragmentation of the BEC\,\cite{Stickney_2002,Gaaloul_2007}, but also affect the overall coherence of the system\,\cite{Chen_2003}.

\begin{figure}[!t]
\begin{center}
\resizebox{0.9\columnwidth}{!}{\includegraphics[clip]{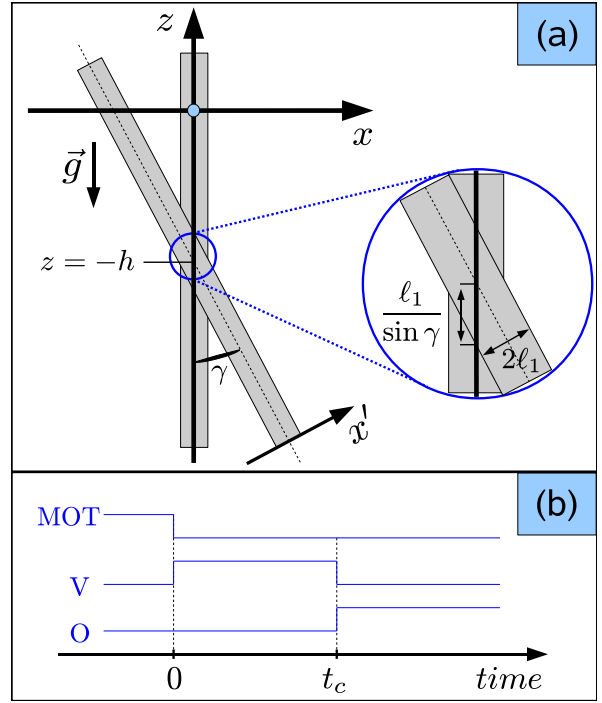}}
\end{center}
\caption{(a) Schematic representation of the proposed optical deflector for cold atoms. The right inset is a magnification of the crossing region. The vertical position of the crossing point is $z=-h$, and the total transverse width of the oblique guide is equal to $2\ell_1$. (b) Timing of the magnetic and optical trap (MOT) and of the vertical (V) and oblique (O) guides used in this setup. $t_c=\sqrt{2h/g}$ corresponds to the time at which the Rb atoms reach the crossing height $z=-h$.}
\label{fig:f1}
\end{figure}

In a recent paper we have derived a semi-classical mo\-del for the description of the splitting dynamics of a cold atomic cloud in such a device\,\cite{Gaaloul_2006}. This setup involves two crossing far off-resonant dipole guides [see Fig.\,\ref{fig:f1}(a)], and we have shown that a simple variation of the laser beam intensities allows to control the splitting ratio in the two guides. In the present paper, we first show that if the vertical guide is switched off when the atomic cloud reaches the crossing point, this device becomes an efficient coherent atom deflector. We then extend this study to the quantum degenerate regime, in order to demonstrate the efficiency of this deflection setup with Bose-Einstein condensates.

As illustrated in Fig.\,\ref{fig:f1}(a), we use a setup involving two crossing far off-resonant dipole guides similar to the one of Ref.\,\cite{Houde_2000}. A large ensemble of $^{87}$Rb atoms is initially trapped and cooled around the position $z=0$ in Fig.\,\ref{fig:f1}(a). This trap is switched off at time $t=0$, while a vertical far off-resonant laser beam, crossing the cloud close to its center, is switched on. A significant portion of the atoms, falling due to gravity, is captured and guided in this vertical wave-guide\,\cite{Houde_2000}. When the center of the guided cloud reaches a given height $z=-h$, at time $t=t_c$, the vertical laser beam is switched off while a second oblique guide is switched on. This timing sequence is illustrated schematically in Fig.\,\ref{fig:f1}(b). The durations of the switching-on and -off procedures are supposed to be much shorter than the typical time scale of the fall dynamics. In spite of the high velocities achieved in this vertical fall, we will show that this setup allows for the implementation of an efficient deflector since the atoms can be deviated from their initial trajectory with no significant loss. This scheme is used both with a thermal cloud of atoms and with an atomic condensate after rescaling the whole problem due to the difference in size of condensates compared to cold atomic clouds.

The outline of the paper is as follows: in Sec.\,\ref{sec:Pot} we discuss the properties of $^{87}$Rb atoms that are relevant for our analysis.  We also give the values of typical laser parameters that realize this atom deflector. We describe briefly our semi-classical numerical model in Sec.\,\ref{sec:Model-cold}. In Sec.\,\ref{sec:Results-cold} we give the results of our numerical investigations on the performance of this setup with cold atomic clouds ($T\sim 10\,\mu$K). We show that a high efficiency ($\geqslant 90\%$) can be achieved with large deflection angles. We also discuss the adiabaticity of the deflection process. We then present in Sec.\,\ref{sec:Bec} a full quantum model designed to treat the dynamics of a BEC falling in the gravity field in the presence of these time-dependent guiding potentials. We then present the results of the numerical simulations with BECs, demonstrating the efficiency of the proposed setup in the quantum degenerate domain. Our conclusions are finally summarized in the last section.

\section{Guiding potentials}
\label{sec:Pot}

During the guiding process and in the case of a large detuning, the atoms are subjected to a dipole force induced by the dipole potential
\begin{equation}
\label{eq:pot}
{\cal U}({\textbf{\em r}}) = \frac{\hbar\Gamma}{2}\,\frac{I({\textbf{\em r}})/I_s}{4\delta/\Gamma}\,,
\end{equation}
where $\delta=\omega_L-\omega_0$ denotes the detuning between the laser frequency $\omega_L$ and the atomic transition frequency $\omega_0$. $I_s$ is the saturation intensity, and $\Gamma$ the natural linewidth of the atomic transition\,\cite{Phillips_1992,Grimm_2000}.

The atomic dynamics is supposed to take place in the $(x,z)$ plane defined by the two guides (see Fig.\,\ref{fig:f1}(a)) thanks to a strong confinement applied in the $y$-direction. The transverse intensity distribution of the TEM$_{00}$ vertical laser beam of power $P_0$ is approximated by the Gaus\-sian-like form
\begin{equation}
\label{eq:int}
\begin{array}{lccl}
\textrm{if }|x| \leqslant \ell_0\textrm{ : } & I_0(x) & = & \displaystyle\frac{2P_0}{\pi w_0^2}\,\sin^2\left(\frac{\pi}{2}\,\frac{x-\ell_0}{\ell_0}\right)\,,\\
\textrm{if }|x| >         \ell_0\textrm{ : } & I_0(x) & = & 0\,,
\end{array}
\end{equation}
where the size $\ell_0$ of the vertical guide is simply related to the laser waist $w_0$ by the relation
\begin{equation}
\label{eq:size}
\ell_0 = w_0 \sqrt{2\ln2} \sim 1.18\,w_0\,.
\end{equation}
This sinus-squared shape, which is often used in time-dependent calculations\,\cite{Giusti-Suzor_1995}, is very close to the ideal Gaussian intensity distribution, except for the absence of the extended wings of the true Gaussian shape which lengthen the calculations without noticeable contribution to the physical processes. With this sinus-squared convention, the guiding region ($|x| \leqslant \ell_0$) is also well defined. The trapping potentials associated with the vertical and oblique laser guides are thus expressed as
\begin{subequations}
\label{eq:U0U1}
\begin{eqnarray}
\label{eq:U0}
{\cal U}_0(x)   & = & -U_{0} \sin^{2}\left(\frac{\pi}{2}\,\frac{x -\ell_0}{\ell_0}\right)\textrm{~~for }|x| \leqslant \ell_0\\
\label{eq:U1}
{\cal U}_1(x,z) & = & -U_{1} \sin^{2}\left(\frac{\pi}{2}\,\frac{x'-\ell_1}{\ell_1}\right)\textrm{ for }|x'| \leqslant \ell_1
\end{eqnarray}
\end{subequations}
where $x'$ denotes the rotated coordinate $x' = x\cos\gamma+(z+h)\sin\gamma$.

Typical laser powers $P_0 \sim 5-30$\,W for a Nd:YAG laser operating at 1064\,nm with laser waists of about $100-300\,\mu$m yield potential depths of about $5-250$\,$\mu$K. With these laser parameters, the $^{87}$Rb transition to consider is the D$_1$\,: 5$^2$S$_{1/2}\rightarrow$\,5$^2$P$_{1/2}$, with a decay rate $\Gamma/2\pi \simeq 5.75$\,MHz, a saturation intensity $I_s \simeq 4.5$\,mW/cm$^2$ and a detuning $\delta/2\pi \simeq -95.4$\,THz. With these conditions, the Rayleigh range $z_R = \pi w_0^2 / \lambda$ is about 3\,cm, thus allowing us to neglect the divergence of the beam on a length up to about 1\,cm.

\section{Semi-classical model for cold atoms}
\label{sec:Model-cold}

The guided atomic dynamics can be followed by solving numerically the time-dependent Schr\"{o}dinger equation for the atomic translational coordinates, taking into account the effect of the gravity field, and choosing realistic values for all laser parameters. We adopt a semi-classical approach where the $z$ coordinate is described classically, following
\begin{subequations}
\label{eq:z(t)}
\begin{eqnarray}
t \leqslant t_c :\;\; z_{cl}(t) & = & -gt^2/2\\
t > t_c         :\;\; z_{cl}(t) & = & -g\big[t_c+(t-tc)\cos\gamma\big]^2/2\,,
\end{eqnarray}
\end{subequations}
where $t_c=(2h/g)^{1/2}$ is the time at which the atoms reach the crossing point (position \mbox{$z=-h$}). These equations of motion are obtained under the assumption of energy conservation for a classical particle which is perfectly deflected, and which therefore follows the paths blazed initially by the vertical beam and later on by the oblique guide. The other dimension $x$ is treated at the quantum level. This semi-classical approach was compared to the experimental study\,\cite{Houde_2000} in Ref.\,\cite{Gaaloul_2006}. In this approach, the two-dimensional guiding potentials\,(\ref{eq:U0U1}) can be replaced by the one-dimensional time-dependent potential
\begin{subequations}
\label{eq:U(x,t)}
\begin{eqnarray}
t \leqslant t_c :\;\; {\cal U}(x,t) & = & {\cal U}_0(x)\\
t > t_c         :\;\; {\cal U}(x,t) & = & {\cal U}_1(x,z_{cl}(t))\,,
\end{eqnarray}
\end{subequations}
and the quantum dynamics is now summarized in the one-dimensional time-dependent Hamiltonian
\begin{equation}
\label{eq:H(x,t)}
\hat{\mathcal{H}}(x,t) = -\frac{\hbar^{2}}{2m}\frac{\partial^{2}}{\partial x^{2}} + {\cal U}(x,t)\,,
\end{equation}
where $m$ denotes the $^{87}$Rb atomic mass. The time-de\-pen\-dent Schr\"{o}dinger equation
\begin{equation}
\label{eq:TDSE}
i\hbar \frac{\partial }{\partial t}\varphi (x,t)=\hat{\mathcal{H}}(x,t)\;\varphi (x,t)\,,
\end{equation}
is then solved using the numerical split operator technique of the short-time propagator\,\cite{Feit_1983}, assuming that the atom is initially ($t=0$) in a well defined eigenstate $v$, of energy $E_v$, of the potential\,(\ref{eq:U0}) created by the vertical laser beam. In addition, it was shown in Ref.\,\cite{Gaaloul_2006} that the deflection probability obtained for the initial classical conditions $z(0)=0$ and $\dot{z}(0)=0$ is very close to the probability averaged over the entire atomic cloud. We therefore use these initial classical conditions in the present study.

At the end of the propagation, the final wave function $\varphi (x,t_{f})$ is analyzed spatially, in order to extract the deflection efficiency $\eta_D$. An averaging procedure over the set of all possible initial states finally allows to calculate the total deflection probability $\langle\eta_D\rangle$ of the entire atomic cloud (see Sec.\,\ref{sec:multiv} hereafter for details).

\section{Numerical Results for cold atoms}
\label{sec:Results-cold}

\subsection{Case of a single initial state}

In this study, the value of the position $h$ of the crossing point between the two guides is the main parameter which controls the efficiency of the deflector. Indeed, for large values of $h$ the atoms reach the crossing point with a large kinetic energy $E_c=mgh$, and they will not be deflected if this energy exceeds by far the binding energy in the oblique guide.

\begin{figure}[!t]
\begin{center}
\resizebox{0.9\columnwidth}{!}{\includegraphics[clip]{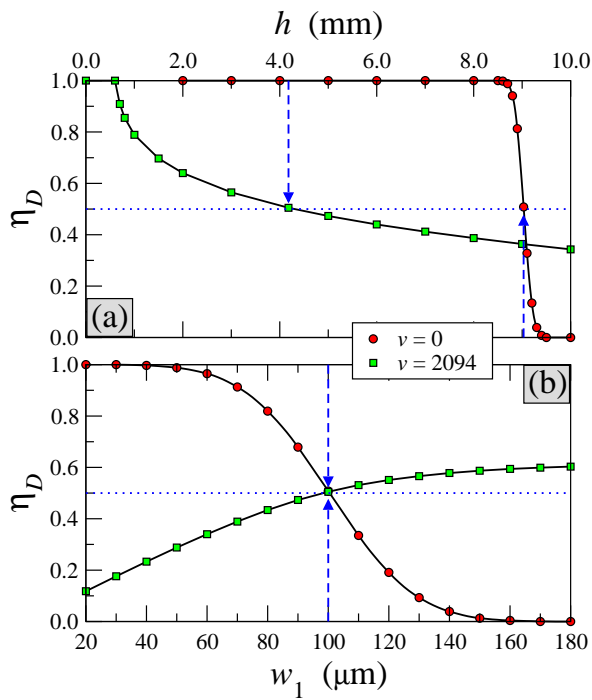}}
\end{center}
\caption{Deflection probability $\eta_D$ as a function (a) of the falling distance $h$ [see Fig.\,\ref{fig:f1}] and (b) of the waist $w_1$ of the oblique laser beam. The deflection angle is equal to $\gamma = 10\,$deg. These results are for a single initial state\,: $v=0$ (solid line with red circles) or $v=2094$ (solid line with green squares). In both graphs, the dashed blue arrows mark the positions at which a deflection efficiency of 50\% is expected according to inequality\,(\ref{eq:critere}). The laser parameters have been chosen such that $U_0=U_1=30\,\mu$K, and $w_0=100\,\mu$m. This corresponds to $P/\delta \simeq 3.1 \times 10^{-4}\,$W/GHz. In graph (a) the oblique laser waist is $w_1=100\,\mu$m and in graph (b) the height $h$ is equal to 9.03\,mm for $v=0$ and to 4.18\,mm for $v=2094$.}
\label{fig:f2}
\end{figure}

In order to predict precisely the largest value of the height $h$ allowing for atomic deflection, one should compare the kinetic energy gained by the atoms along the direction $x'$ transverse to the oblique guide at the position $z=-h-\ell_1/\sin\gamma$ [see Fig.\,\ref{fig:f1}(a)] with the binding energy $U_1-E_v$. The energy $E_v$ denotes here the energy of the initial vibrational state $v$. One can effectively expect that the deflection will fail if
\begin{equation}
\label{eq:critere}
m g \left( h + \frac{\ell_1}{\sin\gamma} \right) \sin^2\gamma > U_1-E_v\,.
\end{equation}
In this expression, the $\sin^2\gamma$ factor originates from the fact that the transverse direction $x'$ of the deflecting beam makes an angle $\gamma$ with the fall direction $z$. The validity of this simple prediction is illustrated in Fig.\,\ref{fig:f2}, which represents the deflection probability $\eta_D$ as a function of $h$ [Fig.\,\ref{fig:f2}(a)] and of $w_1=\ell_1/\sqrt{2\ln2}$ [Fig.\,\ref{fig:f2}(b)], all other parameters being fixed. These probabilities are calculated numerically for the initial state $v=0$ and for $v=2094$, whose energy is about halfway in the optical potential $(E_v \simeq -U_0/2)$. In both graphs, the frontiers defined by the inequality\,(\ref{eq:critere}) are indicated by vertical dashed arrows. By comparison with the ``exact'' value obtained from the solution of the time-dependent Schr\"odinger equation\,(\ref{eq:TDSE}), one can notice that these frontiers correspond to a deflection probability of 50\%. This energy criterion, which simply compares the atomic kinetic energy with the binding energy in the oblique guide, can thus be used safely to predict the efficiency of this setup.

One can also notice in Fig.\,\ref{fig:f2}(a) the different variations of $\eta_D$ with $h$ for $v=0$ and for $v=2094$. The different behavior of these two vibrational levels comes from the fact that $v=0$ is associated with a well localized atomic wavefunction, deeply bound in an almost harmonic potential, while $v=2094$ is entirely delocalized over a large spatial range $|x|\leqslant\ell_0/2$, since its energy is about halfway in the potential. As a consequence, $v=0$ satisfies fully the conditions imposed by the Ehrenfest theorem\,\cite{Ehrenfest_1927} and its evolution can be described classically, while $v=2094$ shows a quantum behavior. For $v=0$, as soon as the inequality\,(\ref{eq:critere}) is satisfied, the deflection probability falls to zero, in agreement with the usual dynamics of a classical particle. On the other hand, the stationnary state $v=2094$ can be seen as a coherent superposition of incoming and outgoing wave packets characterized by a rather broad kinetic energy distribution of width $\Delta E_c \sim U_0/2$. The packet moving in the $+x$ direction will be easily captured by the oblique guide, while the packet moving in the opposite direction easily avoids this wave guide. These two different dynamics are not much affected by the exact value of the falling height $h$, and this explains the very slow variation of $\eta_D$ with $h$ in Fig.\,\ref{fig:f2}(a) for $v=2094$.

The variation of $\eta_D$ with $w_1$ [see Fig.\,\ref{fig:f2}(b)] is also opposite for $v=0$ and $v=2094$. The case $v=0$ can again be interpreted classically\,: when $w_1$ increases, the possibility is open for the atoms to fall from a higher distance $d=(h+\ell_1/\sin\gamma)$, thus gaining a larger kinetic energy. This explains the decrease of $\eta_D$ with $w_1$ for $v=0$. This variation is just reversed in the case of $v=2094$. Here, the initial wave function is characterized by a large typical size $\Delta x \sim \ell_0$. An efficient deflector can thus only be obtained if the size of the oblique wave guide remains of the order of, or is higher than, this typical size $\ell_0$. Consequently, for $v=2094$, when $w_1$ decreases below $w_0$, the deflection probability decreases, as seen in Fig.\,\ref{fig:f2}(b).

In addition, it is worth noting that, due to the $\sin^2\gamma$ factor in the inequality\,(\ref{eq:critere}), it is possible to induce an efficient deflection of atoms with relatively large kinetic energies using modest laser powers, as long as the angle $\gamma$ remains small. For instance, in the case $v=0$ shown in Fig.\,\ref{fig:f2}(a) for $\gamma = 10\,$deg, an almost perfect deflection is obtained for $h=8.5$\,mm, even though the total kinetic energy of the atom reaches then about $E_c \sim 900\,\mu$K, {\it i.e.} 30 times the depth of the oblique wave guide. A larger deflection angle could be achieved easily and with a very high efficiency by simply adding a succession of several deflection setups, each one inducing a small deflection of about 10\,degrees.

An important issue for the preservation of the coherence properties of an atomic cloud is the adiabaticity of the process. Previous theoretical studies have shown that similar beam splitter setups are able to conserve the coherence of the system even for a thermal distribution of atoms with an average energy far exceeding the level spacing of the transverse confinement\,\cite{Kreutzmann_2004}. This behavior results from the fact that non-adiabatic transitions are induced by the time derivative operator $d/dt$ which does not couple states of opposite parities, thus preventing nearest neighbor transitions\,\cite{Hansel_2001}. In comparison, transitions to other states presenting the same parity as the initial state are also not favored since they involve larger energy differences\,\cite{Zhang_2006}.

\begin{figure}[!t]
\begin{center}
\resizebox{0.9\columnwidth}{!}{\includegraphics[clip]{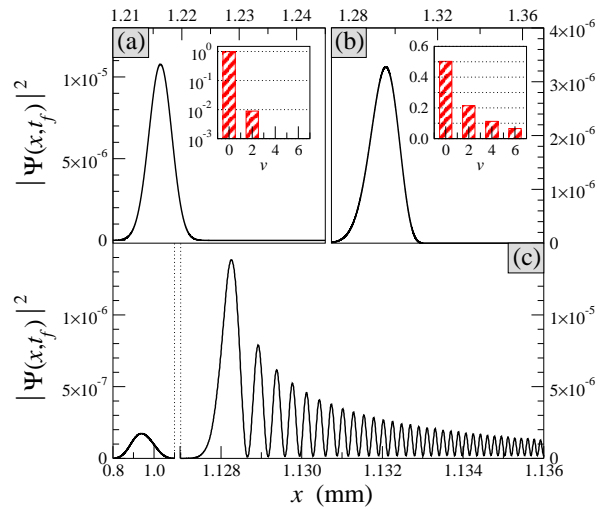}}
\end{center}
\caption{Atomic probability distributions $\left|\varphi (x,t_{f})\right|^2$ as a function of the transverse coordinate $x$ at the end of the propagation, for (a) $h=$2\,mm, (b) $h=$7\,mm and (c) $h=9$\,mm. The laser parameters are identical to the one of Fig.\,\ref{fig:f2}, with $v=0$ and $w_1=100\,\mu$m. Note that, for the sake of clarity, the horizontal axis has been broken in panel\,(c). The small insets in panels (a) and (b) represent the vibrational distributions in the oblique guide at the end of the propagation.}
\label{fig:f3}
\end{figure}

As shown in Fig.\,\ref{fig:f3}, this robustness to non-adiabatic transitions is also present in our deflection scheme. This figure represents the probability distributions $\left|\varphi (x,t_{f})\right|^2$ calculated 7\,mm below the crossing point $z=-h$ for the initial state $v=0$, with $h=2$\,mm [Fig.\,\ref{fig:f3}(a)], $h=7$\,mm [Fig.\,\ref{fig:f3}(b)], and $h=9$\,mm [Fig.\,\ref{fig:f3}(c)]. The vibrational distributions obtained in the oblique guide after deflection are also shown in the small insets of Fig.\,\ref{fig:f3}(a) and\,\ref{fig:f3}(b). Even though the kinetic energy of the atoms exceeds the average vibrational spacing in the trap, the initial state $v=0$ is preserved at 99.1\% for $h=2$\,mm, and at 50.3\% for $h=7$\,mm. Indeed, in the first case, only $v=2$ is slightly populated, while the first five even vibrational levels are populated in the second case. It is only when the falling height $h$ approaches the limit given by the inequality\,(\ref{eq:critere}) that the population of the initial state $v=0$ is almost entirely redistributed to higher excited states, as seen in the wave function shown Fig.\,\ref{fig:f3}(c).

\subsection{Case of an initial vibrational distribution}
\label{sec:multiv}

Realistically, an atomic cloud of typical size $\sigma_0$ and temperature $T_0$ can be described as a statistical mixture of trapped vibrational states, represented by the density matrix
\begin{equation}
\label{eq:thermal}
\rho(\sigma_0,T_0) = \sum_{v} c_v(\sigma_0,T_0) \; | v \rangle \langle v |\,,
\end{equation}
where the coefficients $c_v(\sigma_0,T_0)$ are involved functions of the cloud parameters $\sigma_0$ and $T_0$ and of the wave guide parameters $U_0$ and $w_0$ (see equation (16) in reference\,\cite{Gaaloul_2006} for instance). The calculation of the total deflection probability of the entire cloud therefore requires to average incoherently the deflection probability of each possible initial vibrational level $v$, taking into account the weight functions $c_v(\sigma_0,T_0)$. It is also worth noting that typical initial vibrational distributions $P(v)=|c_v(\sigma_0,T_0)|^2$ are relatively flat when $k_B T \sim U_0$, except for the lowest energy levels which are usually more populated\,\cite{Gaaloul_2006}. In the calculation, we include all populated vibrational states.

\begin{figure}[!t]
\begin{center}
\resizebox{0.9\columnwidth}{!}{\includegraphics[clip]{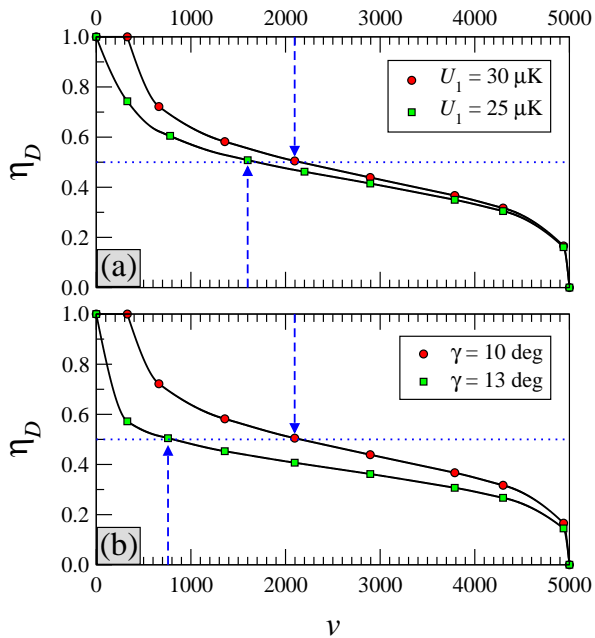}}
\end{center}
\caption{Deflection probability $\eta_D$ as a function of the initial vibrational level $v$. (a) The deflection angle is equal to $\gamma=10\,$deg, and the solid line with red circles stands for $U_1=30\,\mu$K while the solid line with green squares is for $U_1=25\,\mu$K. (b) The depth of the oblique wave guide is equal to $U_1=30\,\mu$K, and the solid line with red circles stands for $\gamma=10\,$deg while the solid line with green squares is for $\gamma=13\,$deg. The falling height is $h=4$\,mm and the oblique laser waist is equal to $w_1=100\,\mu$m. All other parameters are identical to the one of Fig.\,\ref{fig:f2}.}
\label{fig:f4}
\end{figure}

Fig.\,\ref{fig:f4} shows the variation of the deflection efficiency with the initial vibrational level $v$ for a series of different laser parameters. The transverse trapping potential associated with the vertical wave guide supports about 5000 vibrational states when $U_0=30\,\mu$K and $w_0=100\,\mu$m. One can notice the general tendency of measuring a lower deflection probability when $v$ increases, in perfect agreement with the variation expected from the energy criterion\,(\ref{eq:critere}). In addition, one can notice that increasing $U_1$ [Fig.\,\ref{fig:f4}(a)] or decreasing $\gamma$ [Fig.\,\ref{fig:f4}(b)] increases the deflection probability of any initial state. In Fig.\,\ref{fig:f4}, the vertical dashed arrows indicate the limits defined by the inequality\,(\ref{eq:critere}), which are again in good agreement with the numerical values. One can also notice that the highest levels $v \simeq 5000$ are not deflected. This is due to the fact that atoms trapped in these levels, whose energies are very close to the threshold, are easily lost during the deflection process.

\begin{figure}[!t]
\begin{center}
\resizebox{0.9\columnwidth}{!}{\includegraphics[clip]{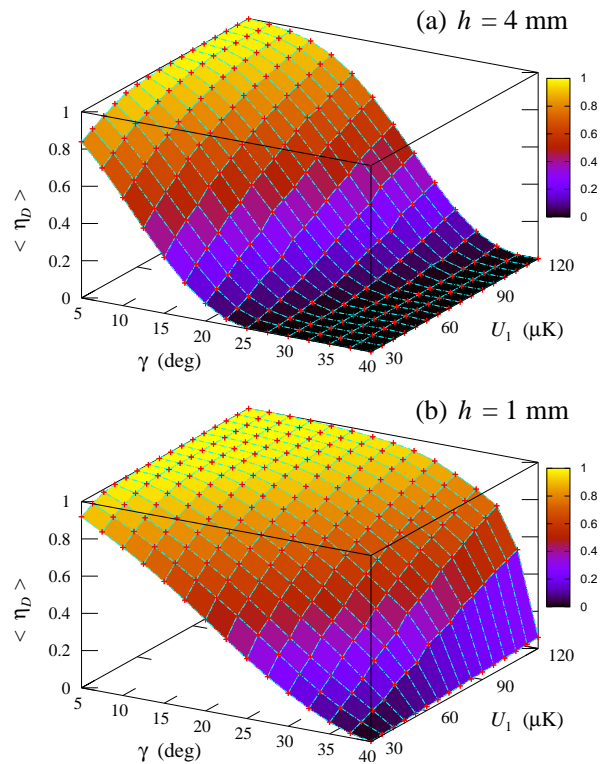}}
\end{center}
\caption{Total deflection probability $\left<\eta_D\right>$ of an atomic cloud of $^{87}$Rb of size $\sigma_0=0.15$\,mm at temperature $T_0=10\,\mu$K. The laser parameters  have been chosen such that $U_0=30\,\mu$K, $w_0=200\,\mu$m, $w_1=158\,\mu$m, and $h=4$\,mm (a) or $h=1$\,mm (b).}
\label{fig:f5}
\end{figure}

Fig.\,\ref{fig:f5} represents the averaged deflection probability $\langle \eta_D \rangle$ as a function of the deflection angle $\gamma$ and of the potential depth $U_1$ of the oblique laser guide, for a thermal input state of size $\sigma_0=0.15\,$mm and temperature $T_0=10\mu$K, with $h=4\,$mm [Fig.\,\ref{fig:f5}(a)] and $h=1\,$mm [Fig.\,\ref{fig:f5}(b)]. Realistic values have been chosen for all laser parameters, close to the one used in the experimental study\,\cite{Houde_2000}, and the coefficients $c_v(\sigma_0,T_0)$ of Eq.\,(\ref{eq:thermal}) were calculated following Ref.\,\cite{Gaaloul_2006}. One can notice a rapid decrease of $\langle \eta_D \rangle$ when $U_1$ decreases and when $\gamma$ increases. However, an almost complete deflection (93.8\%) is still observed in the case $h=1\,$mm with $\gamma=25\,$deg and $U_1=120\,\mu$K, even though the total kinetic energy of the atoms reaches then about $E_c \sim 100\,\mu$K at the crossing point, all trapped states being significantly populated initially. For $\gamma=10\,$deg, the deflection efficiency reaches 99.8\%. We have also verified that decreasing the temperature of the initial cloud increases significantly the deflection efficiency since it suppresses the population of the highest trapped states, for which the deflection process is less efficient [see Fig.\,\ref{fig:f4}]. It is also worth noting that since the deflection process is less efficient for the highest trapped levels, it could also be used to selectively separate the lowest energy levels of the trap. Since it behaves very well for the lowest trapped states, we expect that this setup will prove useful with Bose-Einstein condensates. We therefore derive in the next section a quantum model aimed at the description of the dynamics of a Bose gas in such a deflection setup.

\section{Deflection of Bose-Einstein condensates}
\label{sec:Bec}

\subsection{Theoretical model}
\label{sec:Theo-BEC}

From the theoretical point of view, in the case of a low density the dynamics of the macroscopic wave function $\Psi(\mathbi{r},t)$ of a Bose-Einstein condensate can be accurately described by the mean-field Gross-Pitaevskii equation\,\cite{Gross_1961,Pitaevskii_1961,Gross_1963}. In three dimensions and in the presence of both a time-dependent external potential $V(\mathbi{r},t)$ and the gravity field this equation reads
\begin{equation}
\label{eq:3DGPE}
i\hbar\frac{\partial\Psi}{\partial t} = \Big[ -\frac{\hbar^2}{2m} \nabla^2_{\!r} + V(\mathbi{r},t) + mgz + NU_0\left|\Psi\right|^2 \Big]\Psi\,,
\end{equation}
where $U_0=4\pi\hbar^2a_0/m$ is the scattering amplitude and $a_0$ the $s$-wave scattering length. $N$ denotes the condensate number and $N U_0 \left|\Psi(\mathbi{r},t)\right|^2$ is the mean field interaction energy. The three-dimensional coordinate is denoted by \mbox{$\mathbi{r} \equiv (x,y,z)$}.

In the absence of a trapping potential in the $z$-di\-rec\-tion, the condensate will not only expand but also fall around the average classical height $z_{cl}(t)=-gt^2/2$. Since the de Broglie wavelength of the BEC is no more negligible compared to the characteristic distances of the problem, a quantum treatment of this direction is necessary unlike the thermal atoms case discussed in the first part of the paper. The simulation of the fall dynamics is thus greatly facilitated when done in the moving frame \mbox{$\mathbi{R} \equiv (X,Y,Z)$}, where
\begin{equation}
\label{eq:frame}
\mathbi{R}=\mathbi{r}+\frac{1}{2}\,gt^2\,\mathbi{u}_z\,,
\end{equation}
using the unitary transformation
\begin{equation}
\label{eq:tranf}
\Xi(\mathbi{R},t) = \exp\left[\,i\;\frac{mgt}{\hbar}\left(z+\frac{gt^2}{6}\right)\right] \Psi(\mathbi{r},t)\,.
\end{equation}
Indeed, applying this transformation yields a simplified Gross-Pitaevskii equation
\begin{equation}
\label{eq:3DGPEs}
i\hbar\frac{\partial\Xi}{\partial t} = \Big[ -\frac{\hbar^2}{2m} \nabla^2_{\!R} + V(\mathbi{R},t) + NU_0\left|\Xi\right|^2 \Big] \Xi\,,
\end{equation}
where the gravitational term $mgz$ has vanished.

Following the variational approach of Ref.\,\cite{Salasnich_2002}, we now assume (as in Sec. \ref{sec:Model-cold} and \ref{sec:Results-cold}) a strong harmonic confinement in the perpendicular $Y$-direction, with
\begin{equation}
\label{eq:poty}
V(\mathbi{R},t) = \frac{1}{2} m \omega_\perp^2 Y^2 + V_\parallel(X,Z,t)\,,
\end{equation}
where $V_\parallel(X,Z,t)$ denotes the optical guiding potential
\begin{subequations}
\label{eq:potpar}
\begin{eqnarray}
t \leqslant t_c :\;\; V_\parallel(X,Z,t) & = & {\cal U}_0(X)\\
t > t_c         :\;\; V_\parallel(X,Z,t) & = & {\cal U}_1(X,Z-gt^2/2)\,.
\end{eqnarray}
\end{subequations}
The confinement along the perpendicular direction $Y$ is supposed to be much stronger than along the parallel directions $X$ and $Z$, thus yielding the conditions
\begin{equation}
\label{eq:confine}
\omega_\perp \gg \left[\frac{4U_{0}}{mw_{0}^2}\right]^{\frac{1}{2}}
\quad\textrm{and}\quad
\omega_\perp \gg \left[\frac{4U_{1}}{mw_{1}^2}\right]^{\frac{1}{2}}\,.
\end{equation}
The condensate dynamics is now followed using the appropriate ansatz\,\cite{Salasnich_2002,Jackson_1998}
\begin{equation}
\label{eq:trialwf}
\Xi(\mathbi{R},t) = \Phi(X,Z,t)\;f\big(Y|\Omega\big)\,,
\end{equation}
where
\begin{equation}
\label{eq:trialwfG}
f\big(Y|\Omega\big) = \frac{e^{-\frac{1}{2}\frac{Y^2}{\Omega^2}}}{\pi^{\frac{1}{4}}\Omega^{\frac{1}{2}}}\,.
\end{equation}
This choice amounts to assume a Gaussian shape of the wave function in the $Y$-direction, characterized by a time-dependent width $\Omega(X,Z,t)$. This width varies slowly along the  parallel directions, thus implying
\begin{equation}
\nabla^2_{\!R} f \simeq \partial^2 f/\partial Y^2\,.
\end{equation}
It has been shown that this choice is well justified not only in the limit of weak interatomic couplings but also with large condensate numbers\,\cite{Perez-Garcia_1996,Perez-Garcia_1997,Parola_1998}.

An effective two-dimensional non-linear wave equation is then derived using the quantum least action principle\,\cite{Schiff_1968,Salasnich_2002} for $\Phi(X,Z,t)$
\begin{eqnarray}
\label{eq:2DGPE}
i\hbar\frac{\partial\Phi}{\partial t} & = & \Big[ -\frac{\hbar^2}{2m} \nabla^2_{\parallel} + V_\parallel +
                                               \frac{N\,U_0}{\sqrt{2\pi}}\,\frac{\left|\Phi\right|^2}{\Omega}\nonumber\\
                                      &   & \qquad + \frac{1}{4}\left(\frac{\hbar^2}{m\Omega^2} + m\omega_\perp^2\Omega^2\right) \Big] \Phi\,.
\end{eqnarray}
This equation describes the condensate dynamics in the $(X,Z)$ plane, with an accuracy which goes beyond the usual two-dimensional Gross-Pitaevskii equation. It takes into account the influence of the dynamics along the perpendicular direction on the evolution of $\Phi(X,Z,t)$ with the introduction of the width parameter $\Omega(X,Z,t)$. Note the difference by a factor $1/2$ in the last two terms of Eq.(\ref{eq:2DGPE}) when compared to Eq.(25) of Ref.\,\cite{Salasnich_2002}, due to a misprint in Ref.\,\cite{Salasnich_2002}. The least action variational principle also yields the following quartic equation governing the evolution of this parameter
\begin{equation}
\label{eq:Omega}
\left(\frac{1}{2}m\omega_\perp^2-\frac{2V_\parallel}{w^2(t)}\right)\Omega^4-\frac{N\,U_0}{2\sqrt{2\pi}}\left|\Phi\right|^2\,\Omega-\frac{\hbar^2}{2m}=0\,,
\end{equation}
where $w(t)=w_0$ for $t \leqslant t_c$ and $w(t)=w_1$ for $t > t_c$. This last equation was obtained assuming $\Omega(X,Z,t) \ll w(t)$ for all $X$, $Z$ and $t$, in agreement with the strong confinement in $Y$. Compared to Eq.(26) of Ref.\,\cite{Salasnich_2002}, the additional term $2V_\parallel/w^2(t)$ is a small correction due to the $Y$-dependence of the TEM$_{00}$ laser intensity profile.

The time-dependent wave equation\,(\ref{eq:2DGPE}) is solved numerically using the splitting technique of the short-time propagator\,\cite{Feit_1983}, while the quartic equation\,(\ref{eq:Omega}) for $\Omega$ is solved at each time step and for each coordinate grid point $(X,Z)$ using an efficient numerical algorithm\,\cite{Hacke_1941}. The initial wave function is obtained using the imaginary time relaxation technique\,\cite{Kosloff_cpl}, and the three-dimensional condensate wave function $\Psi(\mathbi{r},t)$ is reconstructed at the end of the propagation by inverting the transformations\,(\ref{eq:frame}) and\,(\ref{eq:tranf}).

\begin{figure}[!t]
\begin{center}
\resizebox{0.9\columnwidth}{!}{\includegraphics[clip]{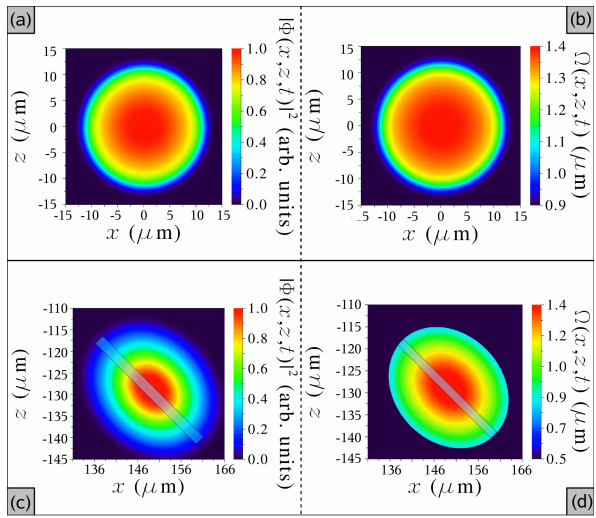}}
\end{center}
\caption{(a) and (c)\,: Atomic density $|\Phi(x,z,t)|^2$ in arbitrary units as a function of $x$ and $z$ at times $t=0$ (upper graph) and $t=5.2\,$ms (lower graph). (b) and (d)\,: Width parameter $\Omega(x,z,t)$ in $\mu$m as a function of $x$ and $z$ at times $t=0$ (upper graph) and $t=5.2\,$ms (lower graph). In each sub-plot, the color scale is defined on the right hand side of the graph. The guiding potentials are defined by $U_0=2.2\,\mu$K [$P/\delta \simeq 2.3 \times 10^{-5}\,$W/GHz], $U_1=8.8\,\mu$K [$P/\delta \simeq 9.2 \times 10^{-5}\,$W/GHz], and $w_0=w_1=300\,\mu$m. The deflection angle is $\gamma=50\,$deg and the condensate number is $N=5 \times 10^4$. In the $y-$direction, the trapping frequency $\omega_y$ is assumed to be 10 times larger than $\omega_x$. The crossing height is $z=-h=-10\,\mu$m and the time step for the split operator numerical propagation is $\delta t=1\,\mu$s.}
\label{fig:f6}
\end{figure}

\subsection{Numerical Results}
\label{sec:Res-BEC}

A typical BEC dynamics is illustrated in Fig.\,\ref{fig:f6}, which shows a surface plot of the atomic density $|\Phi(x,z,t)|^2$ and of the width parameter $\Omega(x,z,t)$ at time $t=0$ [upper part, labels (a) and (b)] and during the deflection process, at time $t=5.2\,$ms [lower part, labels (c) and (d)]. In this numerical example, the condensate number is $N=5 \times 10^4$. The crossing height $z=-h$ is reached at time $t_c=1.43\,$ms, and the deflection angle is fixed at the value $\gamma=50\,$deg. In the lower graphs the transparent oblique line indicates the direction of deflection corresponding to this angle.

\begin{figure}[!t]
\begin{center}
\resizebox{0.9\columnwidth}{!}{\includegraphics[clip]{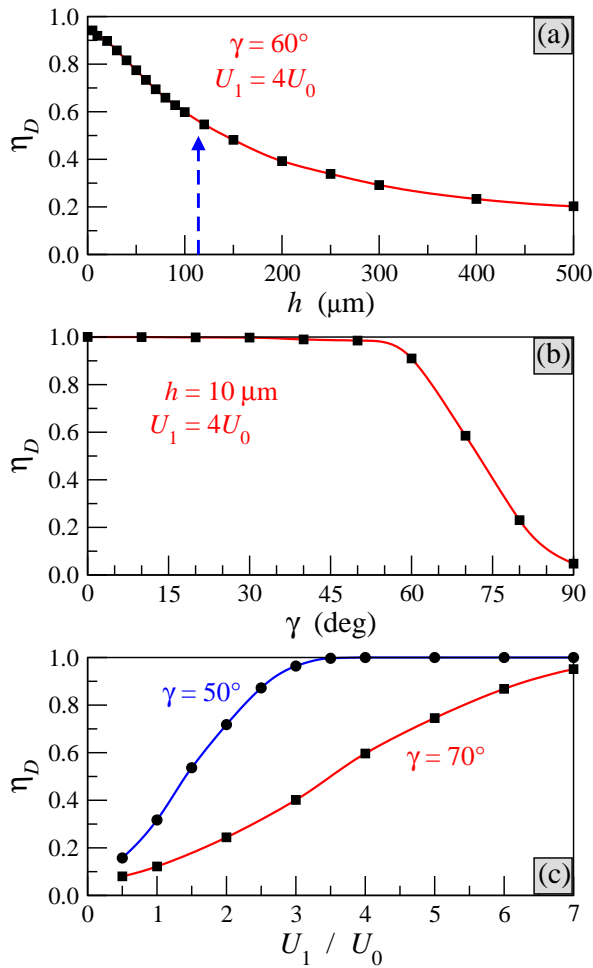}}
\end{center}
\caption{Condensate deflection efficiency $\eta_D$ as a function of (a) the crossing height $h$, (b) the deflection angle $\gamma$ and (c) the ratio of binding energies $U_1/U_0$, equivalent to the ratio of laser powers $P_1/P_0$. In these three graphs, the potential depth $U_0$ is fixed at 2.2\,$\mu$K, and $w_0=w_1=300\,\mu$m. The condensate number is $N=5 \times 10^4$. In (a) $U_1=8.8\,\mu$K and $\gamma=60\,$deg. In (b) $U_1=8.8\,\mu$K and $h=10\,\mu$m. In (c) $h=10\,\mu$m and $\gamma=50\,$deg (blue line with circles) and $\gamma=70\,$deg (red line with squares). The dashed blue arrow in the upper graph marks the position at which a deflection efficiency of 50\,\% is expected according to Eq.(\ref{eq:critere}).}
\label{fig:f7}
\end{figure}

One can first notice in Fig.\,\ref{fig:f6}(b) and (d) that even for a relatively small condensate number, a Gaussian ansatz for $f(Y|\Omega)$ can only be used if the width $\Omega$ is a free parameter which can take different values at different positions $x$ and $z$. This result is in agreement with the numerical studies of  Salasnich \emph{et al}\,\cite{Salasnich_2002}.

The inset (a) of Fig.\,\ref{fig:f6} shows the symmetric ground state wave function of the initial condensate prepared in a parabolic trapping potential of equal frequencies $\omega_x=\omega_z$. At time $t=5.2\,$ms [insets (c) and (d) of Fig.\,\ref{fig:f6}], the condensate wave function has expanded during the fall dynamics, and has been efficiently deflected along the direction $\gamma=50\,$deg by the oblique laser guide.

In our numerical simulations, we propagate the condensate wave function well after the crossing point between the vertical and oblique laser beams has been rea\-ched, and we obtain the deflection efficiency $\eta_D$ by calculating the condensate number in the oblique trapping potential at the end of the propagation. Fig.\,\ref{fig:f7} shows the variation of the condensate deflection efficiency with the crossing height $h$ [inset (a)], the deflection angle $\gamma$ [inset (b)] and the ratio of laser powers $P_1/P_0=U_1/U_0$ [inset (c)]. The other parameters are given in the figure caption.

The simple energy criterium\,(\ref{eq:critere}) is still shown to be relatively accurate with Bose-Einstein condensates since in Fig.\,\ref{fig:f7}(a), a deflection efficiency of 50\,\% is expected for a falling height $h = 114\,\mu$m according to Eq.(\ref{eq:critere}), and the numerical simulation yields a deflection probability of 55\,\%. The variation of the deflection efficiency $\eta_D$ with $h$, $\gamma$ and $U_1/U_0$ is also found to be very similar to the one obtained with cold atomic clouds. One can finally notice that large deflection angles can be reached when $U_1 > U_0$, with almost no atom loss for instance when $\gamma = 50\,$deg and $U_1 = 4 U_0$. This high deflection efficiency can be explained by the small size of the BEC and the weak spread in velocities compared to cold atoms.

\section{Conclusion}
\label{sec:Conclusion} 

In summary, we have presented a detailed analysis of the implementation of an optical deflector for cold atomic clouds and for Bose-Einstein condensates. Our analysis is quite close to the experimental conditions, and is clearly within the reach of current technology. We have shown how to create a high performance deflector using two crossing laser beams which are switched on and off in a synchronized way. We have found that a 10\,$\mu$K cloud of Rubidium atoms can be deflected by 25\,degrees with an efficiency of about 94\%, and by 10\,degrees with an efficiency exceeding 99\%. A succession of such deflecting setups at this small angle could also be implemented in order to achieve larger deflection angles with high fidelities. We have shown that this device is robust against non-adiabatic transitions, an undesirable effect which could have led to heating processes. A high degree of control can therefore be achieved with such quantum systems, opening some possibilities for a range of applications. We have also derived an original approach treating the dynamics of a Bose-Eintein condensate in the gravity field using the quantum least action principle in a moving frame. This model was used to demonstrate the high efficiency of this deflection setup with quantum degenerate gases since deflection angles of up to about 50\,degrees can be implemented with no significant atom loss.

\begin{acknowledgements}
We acknowledge HPC facilities of the IDRIS-CNRS supercomputer center (Project No.\,08-51459) and E.C acknowledges partial financial support from CEA via the LRC-DSM Grant No.\,05-33. E.C. and L.P. acknowledge financial support from PPF ``Information Quantique'', and L.P. acknowledges financial support from IFRAF (Institut Francilien de Recherche sur les Atomes Froids). Laboratoire de Photophysique Mol\'eculaire and Laboratoire Aim\'e Cotton are associated to Universit\'e Pa\-ris-Sud.
\end{acknowledgements}

\end{document}